\documentstyle[graphicx]{aipproc}

\begin{document}
\title{
A Fundamental Limit of Measurement Imposed by the 
Elementary Interactions
}

\author{
Akira Shimizu
\thanks{
E-mail: shmz@ASone.c.u-tokyo.ac.jp
}
}
\address{
Department of Basic Science, 
University of Tokyo, 
Komaba, Meguro-ku, Tokyo 153-8902, Japan
}

%
\vspace{-40mm}
\begin{center}
{\small
Proc. 3rd Tohwa Univ. Int. Conf. Statistical Physics 
(Fukuoka, Japan, 1999)

AIP, to be published
}
\end{center}

\maketitle

\begin{abstract}
Quantum information theory is closely related to 
quantum measurement theory because 
one must perform measurement 
to obtain information on a quantum system.
Among many possible limits of quantum measurement, 
the simplest ones were
derived directly from the uncertainty principles.
However, such simple limits are not the only limits. 
I here suggest a new limit which 
comes from the forms and the strengths of the 
elementary interactions.
Namely, there are only four types of elementary interactions 
in nature; their forms are determined by the gauge invariance
(and symmetry breaking), 
and their coupling constants (in the low-energy regime) 
have definite values.
I point out that this 
leads to a new fundamental limit of quantum measurements.
Furthermore, this fundamental limit imposes the fundamental 
limits of getting information on, preparing,   
and controlling quantum systems.
\end{abstract}

\section*{Introduction}

Quantum information theory is closely related to 
quantum measurement theory because 
one must perform measurement 
to obtain information on a quantum system.
In particular, 
a limit of quantum measurement 
is also a limit of getting information on 
a quantum system.
Among many possible limits of quantum measurement, 
the simplest ones were
derived directly from the uncertainty principles.
However, {\it such simple limits are not the only limits}. 
I here suggest a new limit which 
comes from the forms and the strengths of the 
elementary interactions.
%

It is established that 
measurement of a quantum system can be analyzed 
(except for philosophical issues) as 
interaction processes of the system and the measuring apparatus
\cite{LP,QNDmass,QNDphoton,PSY,ozawa,SF,pra,Nano}.
Many interesting quantities, such as a measurement error and 
backaction of the measurement, can be calculated 
from the dynamics of the total quantum system composed of 
the system and the measuring apparatus 
\cite{LP,QNDmass,QNDphoton,PSY,ozawa,SF,pra,Nano}.
The dynamics of a quantum system
is governed by the commutation relations 
{\it and} the Hamiltonian. 
For the interaction Hamiltonian $H_I$, 
it was often {\it assumed} in the literature
\cite{QNDmass,QNDphoton,PSY} that 
the form of $H_I$ was obtained by ``quantizing'' 
some classical Hamiltonian.
Or, in some cases \cite{ozawa}, specific forms were 
{\em just assumed} for $H_I$.
Discussions based on such $H_I$ may be consistent {\it 
mathematically}. Physically, however, 
such $H_I$ can be wrong. 
For example, quantization of a classical Hamiltonian in general 
gives a form that is different from the true Hamiltonian, 
and the difference
is typically of the order of $\hbar$. 
However, backaction is also of the order of $\hbar$. 
Hence, physically, any results 
drawn from such a Hamiltonian cannot be reliable enough. 
In particular, 
{\it there are only four types of elementary interactions 
in nature}; 
their forms are determined by the gauge invariance
(and symmetry breaking), 
and their coupling constants (in the low-energy regime) 
have definite values \cite{ft}.
In order to draw 
conclusions which are physically meaningful, one must either 
use these true interactions, or, if one uses an approximate Hamiltonian, 
one must confirm that the approximation does not alter the results 
for the measurement. 
Unfortunately, however, these points were not examined carefully.

In this paper, I point out that 
not only the commutation relations
but also the limitation of available interactions 
in nature results in a fundamental limit of 
quantum measurement.
This fundamental limit imposes the fundamental 
limits of getting information on, preparing,   
and controlling quantum systems.

%
%
%
%

\section*{
Measurement error and backaction on the measured observable}


I consider the relation between
the measurement error and the backaction {\it on the measured observable}.
Although similar arguments should be applicable to other relations, 
I here present only one example, to explain 
the basic ideas.

When one measures an observable $Q$ of a quantum system, 
there is a finite measurement error $\delta Q_{err}$ 
in general.
Its magnitude is determined by 
the form and 
strength of the interaction Hamiltonian $H_I$ between 
the measured system and the measuring apparatus 
\cite{QNDmass,QNDphoton,PSY,ozawa,SF,pra,Nano}. 
There are also backactions of the measurement.
A well-known backaction
 is the one on the conjugate variable $P$ of $Q$.
Its lower limit
is determined by the commutation relation
between $Q$ and $P$, 
i.e.,  by 
the uncertainty principle.
However, I here discuss backaction on $Q$ {\it itself} 
\cite{LP,QNDmass,QNDphoton,PSY,ozawa,SF,pra,Nano}, 
which is denoted by $\delta Q_{ba}$. 
Although the ``ideal measurement'',
for which $\delta Q_{ba} = \delta Q_{err} =0$,
is normally  {\it assumed} in textbook quantum mechanics, 
real measurements are non-ideal in general, 
for which $\delta Q_{ba}$ and/or $\delta Q_{err}$
are finite
\cite{LP,QNDmass,QNDphoton,PSY,ozawa,SF,pra,Nano}.
Since the lower limit of $\delta Q_{ba}$ is not (directly) limited 
by the uncertainty principle, 
$\delta Q_{ba}$ strongly depends on the form and 
strength of the interaction Hamiltonian $H_I$
\cite{LP,QNDmass,QNDphoton,PSY,ozawa,SF,pra,Nano}, just as $\delta Q_{err}$ does.
Therefore, $\delta Q_{ba}$ and $\delta Q_{err}$ are correlated
through $H_I$.
For this relation, one could not get physically correct results 
without using a correct Hamiltonian. 

For example, if one {\it assumes} some $H_I$ which satisfies
\begin{equation}
[H_I, Q] = 0,
\end{equation}
then $\delta Q_{ba}=0$, independent of the strength 
(which may be expressed by the coupling constant $g$) 
of $H_I$ \cite{QNDmass,QNDphoton,SF}.
Namely, the measurement becomes of the first kind (FK)
\cite{LP,SF}.
On the other hand, 
$\delta Q_{err}$ tends to increase as $g$ is decreased:  
For example, $\delta Q_{err} \to \infty$ as $g \to 0$ 
\cite{QNDphoton,PSY,ozawa,SF,pra,Nano}.
Hence, one would conclude that
\begin{equation}
\mbox{$\delta Q_{ba}$ and $\delta Q_{err}$ are uncorrelated.}
\label{wrong}\end{equation}
However, this is true only for such a hypothetical $H_I$.
Namely, if 
\begin{equation}
[H_I, Q] \neq 0
\end{equation}
for the true form of $H_I$ (i.e., for 
the elementary interaction),  
then 
$\delta Q_{ba}$ strongly depends on the form and 
strength of $H_I$.
Therefore, the correct conclusion is
\begin{equation}
\mbox{$\delta Q_{ba}$ and $\delta Q_{err}$ are strongly correlated.}
\label{correct}\end{equation}
in sharp contrast to (\ref{wrong}).

Unfortunately, 
most previous work on measurement
\cite{QNDmass,QNDphoton,PSY,ozawa}
assumed some ``effective'' interactions for $H_I$, 
such as a photon-photon interaction, 
which is not the true form,  
the elementary interaction.
The validity of such  effective forms,  
when applied to problems of measurement, was not 
examined carefully.

\section*{
measurement of the photon number}


From this viewpoint, of particular interest is the measurement of a 
gauge field such as the photon field. 
The form of its interaction 
is completely determined 
by the requirement of the gauge invariance:
the only interaction is the gauge-invariant interaction 
with the matter field \cite{ft}.
However, most previous work on measurement of a gauge field, 
particularly work on the 
``quantum non-demolition (QND) measurement''
of photons \cite{QNDphoton,PSY}, 
assumed other forms for $H_I$,  
such as a photon-photon interaction,
as ``effective'' interactions.
Here, the QND measurement is the FK measurement for 
a conserved observable, i.e., for $Q$ 
that is conserved during the free motion (i.e., while 
the system is decoupled from the measuring apparatus)
\cite{QNDmass}.
It therefore seems that most previous discussions on measurement 
of a gauge field need to be reconsidered.

\begin{figure}[h]
\begin{center}
\includegraphics[width=.9\textwidth]{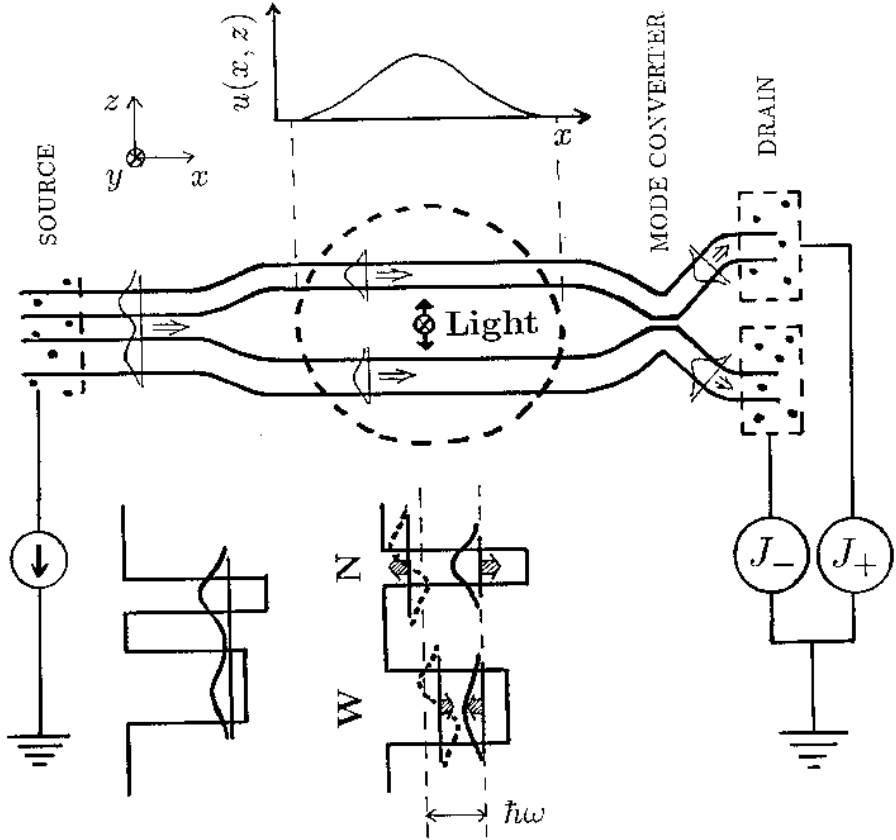}
\end{center}
\caption[]{
A quantum non-demolition
photodetector composed of a double-quantum wire
electron interferometer.
(Taken from \cite{pra})
}
\label{fig1}
\end{figure}

To demonstrate the dramatic difference between 
the correct result derived from a proper $H_I$ 
and a wrong result derived from an ``effective'' $H_I$, 
I will present results for 
the QND photodetector proposed in Refs.\ [3,4], 
for both cases where the true $H_I$ and 
an ``effective'' $H_I$ are used.
%
 
A schematic diagram of the
QND photodetector
is shown in Fig.1 \cite{pra}. 
Before going to the full analysis in the following
sections, I here give an intuitive, semi-classical  description
of the operation principle \cite{SFOY}.
The device is
composed of two quantum wires, N and W, which 
constitute an electron interferometer.
The lowest subband energies (of the $z$-direction confinement)
$\epsilon_a^N$ and $\epsilon_a^W$
of the wires are the same, but the second levels
$\epsilon_b^N$ and $\epsilon_b^W$ are different.
Electrons occupy the lowest levels only.
A $z$-polarized light beam
hits the dotted region. The photon energy $\hbar \omega$ is assumed to
satisfy
$\epsilon_b^W - \epsilon_a^W < \hbar \omega < \epsilon_b^N - \epsilon_a^N$,
so that real excitation does not occur and no photons are absorbed.
However,
the electrons are excited ``virtually" \cite{virtual}, and
the electron wavefunction undergoes a phase shift
between its amplitudes in the two wires.
Since the magnitude of the virtual excitation is proportional to
the light intensity \cite{virtual}, so is the phase shift.
This phase shift modulates the interference currents, $J_+$ and $J_-$.
By measuring $J_\pm$, we can know the magnitude of the phase shift,
from which we can know the light intensity.
Since the light intensity is proportional to the photon number $n$,
we can get information on $n$.
We thus get to know $n$
without photon absorption, i.e.,
without changing $n$; hence the name QND.
(More precisely, 
the distribution of $n$ over the whole ensemble is 
unchanged, 
whereas $n$ of each element of the ensemble 
is changed due to the ``reduction'' of the wavefunction.
See, e.g., section 7 of Ref.\ \cite{Nano}.)
In contrast, conventional photodetectors drastically alter the photon number by
absorbing photons.
Keeping this semi-classical argument in mind,
let us proceed to a fully-quantum analysis.

\section*{Results derived from an ``effective'' Hamiltonian}

As explained above, the essential role of the photon 
field is to modulate the electron phase.
It is thus tempting to employ the following form 
\begin{equation}
H_I = 
\sum_{\nu=N,W} \sum_k g_{k \nu} c_{k \nu}^\dagger c_{k \nu} n,
\label{Heff}\end{equation}
as an ``effective'' interaction Hamiltonian.
Here, $c_{k \nu}$ denotes the annihilation operator
of an electron of wavelength $k$ in quantum wire $\nu$, 
and $g_{k \nu}$ is an effective coupling constant.
This Hamiltonian commutes with the photon number $n$;
\begin{equation}
[H_I, n] = 0.
\label{commute}\end{equation}
This equation, and 
the fact that $n$ is conserved during the free evolution,
meet 
the condition for QND detectors that were claimed in Refs. 
\cite{QNDmass,QNDphoton}.

It is then easy to show that 
the backaction $\delta n_{ba} = 0$, 
whereas 
the measurement error $\delta n_{err}$ is finite
and dependent on $g_{k \nu}$.
Namely, we obtain conclusion (\ref{wrong})
for the interaction Hamiltonian (\ref{Heff}).

\section*{Results derived from the correct Hamiltonian}

The correct interaction (i.e., the elementary interaction
of, in this case, quantum electrodynamics)
is not the form of Eq.\ (\ref{Heff}), 
but the following one \cite{ft};
\begin{equation}
H_I
=
\frac{e}{\hbar c} 
\int d^4 x \
\bar{\psi} \gamma^\mu A_\mu \psi.
\label{qed}\end{equation}
Here, 
$\psi$ is the (relativistic) electron field, 
$\gamma^\mu$ ($\mu = 0, 1, 2, 3$) are the gamma 
matrices, 
$A_\mu$ is the electromagnetic potential
(the set of the vector potential and scalar potential),
and 
$\bar{\psi} = \psi^\dagger \gamma^0$.
This form is uniquely determined by the local gauge 
invariance \cite{ft}.
Furthermore, the value of the coupling constant $e/\hbar c$ 
(more precisely, the renormalized value at 
the low energy region) is uniquely determined by experiment \cite{ft}.
Since $n$ is quadratic in $A_\mu$, we find
\begin{equation}
[H_I, n] \neq 0,
\label{notcommute}\end{equation}
in contrast to Eq.\ (\ref{commute}).
Hence, 
the QND condition claimed in Refs.\ \cite{QNDmass,QNDphoton}
is {\it not} satisfied by the correct Hamiltonian.
According to the detailed analysis on QND measurement \cite{SF}, 
the condition claimed in Refs.\ \cite{QNDmass,QNDphoton} is the strongest 
one among various possibilities, which are 
called the strong, moderate, and weak conditions \cite{SF}.

The weak condition given in Ref.\ \cite{SF} can be slightly 
generalized to the following form \cite{unpub}
(which reduces to that of Ref.\ \cite{SF} as 
$\varepsilon_{ba} \to 0$): 
For a small positive number $\varepsilon_{ba}$,
\begin{equation}
\left|
\sum_{j} 
\left| 
\sum_{k,\ell} a_k b_\ell u_{ij}^{k \ell}
\right|^2
-
|a_i|^2
\right|
\leq
\frac{
\varepsilon_{ba}
}{2}
\left|
\sum_{j} 
\left| 
\sum_{k,\ell} a_k b_\ell u_{ij}^{k \ell}
\right|^2
+
|a_i|^2
\right|
\quad \mbox{for all $i$},
\label{weak}\end{equation}
for some set of $\{ a_k \}$'s and for some $\{ b_\ell \}$.
Here,  
I have used the same notations as in 
Ref.\ \cite{SF}; 
$\{ a_k \}$ and $\{ b_\ell \}$
represent the initial states of the measured system 
and the ``probe'' system, respectively,
i.e., the expansion coefficients 
in terms of the eigenfunctions of the measured and ``readout''
variables \cite{SF}.
The $u_{ij}^{k \ell}$ is the matrix representation of 
the unitary evolution induced by the 
interaction with the probe system,
hence is a function of the Hamiltonian \cite{SF}.
Physically, this condition means that the backaction can be made 
small enough for 
a particular set of measured states
if the measuring apparatus is appropriately designed. 
The magnitude of the backaction is represented by 
$\varepsilon_{ba}$, which gives the upper limit of the 
change of the probability distribution
over the ensemble \cite{SF}.
Note that the limitation of measured states 
 corresponds to a limitation of the response range of the measuring 
apparatus. 
Such a limitation is quite realistic, 
because {\it any} 
existing apparatus do have finite response ranges.
Furthermore, as is clear from Eq.\ (\ref{notcommute}), 
{\it only by accepting such a limitation 
can we realize the FK or QND measurement of wide classes of observables}
including photon number. 

The condition satisfied by the QND photodetector of Fig.\ 1
is this (generalized) weak condition.
In fact, 
the backaction $\delta n_{ba}$ 
which is induced by the QND detector of Fig.\ 1 
is evaluated as \cite{unpub,nonrel}
\begin{equation}
\delta n_{ba}
\propto
{\gamma^2 \langle n \rangle N \over |\Delta|^4 \tau_p},
\label{ba}\end{equation}
where 
$\gamma$ ($\propto e/\hbar c$) denotes a constant that 
is determined by the structures of the quantum 
wires and the optical waveguide,
$N$ is the number of electrons detected as the interference
current, 
$\Delta = \epsilon_b - \epsilon_a - \hbar \omega$ 
is the detuning energy,
and $\tau_p$ denotes the optical-pulse duration
(which is assumed, for simplicity, 
to be longer than $\tau_t$ of Ref.\ \cite{pra}).
It is seen that for a small positive number $\varepsilon_{ba}$, 
one can make
\begin{equation}
\frac{\delta n_{ba}}{\langle n \rangle}
\leq \varepsilon_{ba},
\label{n_to_nba}\end{equation}
by taking $1/|\Delta|$, $1/\tau_p$,  
$\gamma^2$, and/or $N$ small enough.
This corresponds to Eq.\ (\ref{weak}).
Namely, the (generalized) weak condition (\ref{weak}) for the 
QND measurement is indeed satisfied.
In other words, 
the QND measurement is possible
for a certain set of photon states 
(i.e., for photon states whose pulse duration $\tau_p$ is
longer than some value)
if the structure of the device is appropriately 
designed (which determines $\gamma^2$ and $N$).

On the other hand, 
the measurement error $\delta n_{err}$ is evaluated as 
\cite{pra,Nano,nonrel}
\begin{equation}
\delta n_{err} 
\propto 
{|\Delta| \over \gamma^2 \sqrt{N}}.
\label{err}\end{equation}
This dependence on $N$ is a general result for
quantum interference devices \cite{SS}.
It is seen that 
for a small positive number $\varepsilon_{err}$,
one can make
\begin{equation}
\frac{\delta n_{err}}{\langle n \rangle}
\leq \varepsilon_{err},
\end{equation}
by taking 
$1/|\Delta|$, $\gamma^2$, $N$, 
and/or $\langle n \rangle$ 
large enough.
However, as seen from Eq.\ (\ref{ba}), 
their increase 
results in the increase of the backaction.
Therefore, 
the backaction 
(not on the conjugate observable, but on the photon number itself)
and the measurement error 
are 
strongly correlated: 
{\em one 
cannot make $\delta n_{err}$ ($\delta n_{ba}$) smaller without 
accepting the increase of $\delta n_{ba}$ ($\delta n_{err}$).
}
This correct conclusion, 
derived from the correct Hamiltonian (\ref{qed}), 
should be contrasted with the wrong result 
obtained from the ``effective'' Hamiltonian (\ref{Heff}).

\section*{Optimization of the measuring apparatus 
and the upper limit of the information 
entropy}

Because of this unavoidable correlation, 
one must make some optimization.
For example, 
if one wishes to achieve
\begin{equation}
\varepsilon_{ba}, \ \varepsilon_{err} \lesssim 10^{-2}
\quad \mbox{for $\tau_p \geq 10$ ps},  
\label{eps}\end{equation}
then 
the QND detector can be designed 
in such a way that \cite{unpub} 
\begin{eqnarray}
\frac{\delta n_{ba}}{\langle n \rangle}
&\simeq&
10^{-2},
\\
\frac{\delta n_{err}}{\langle n \rangle}
&\simeq&
\frac{10^{2}}{\langle n \rangle}
\quad
\mbox{(i.e., $\delta n_{err} \simeq 10^{2}$)}.
\end{eqnarray}
In this case, Eq.\ (\ref{eps}) 
is satisfied if
\begin{equation}
\langle n \rangle \geq 10^{4}. 
\end{equation}
Hence, 
the lower boundary $n_{min}$ of the response range 
of the QND detector is found to be $n_{min} = 10^{4}$.
%
On the other hand, the upper limit $n_{max}$ of the response range
is evaluated in this case as $n_{max} \sim 10^6$, which 
is derived from the requirement that 
the phase shift of an electron should be less than $\pi$ \cite{pra}.
Then, Eq.\ (\ref{eps}) 
is satisfied for all values of 
$\langle n \rangle$ between $n_{min}$ and $n_{max}$.
Moreover, $\delta n_{err}$ is smaller than the ``standard 
quantum limit'' $\sqrt{\langle n \rangle}$ 
\cite{QNDphoton,PSY}, throughout this range.

It is interesting to evaluate
the information entropy $I$ that can be obtained by the QND detector.
Since $\delta n_{err}$ is 
independent of $\langle n \rangle$, 
the number of different values of 
$\langle n \rangle$ that 
can be distinguished by this QND detector is simply given by
\begin{equation}
\frac{
n_{max} - n_{min}
}{
\delta n_{err}
}
\sim 
10^{4}.
\end{equation}
Hence, $I$ is estimated as
\begin{equation}
I 
\simeq 
\log_2 10^4
\simeq
13.3.
\end{equation}
This should be contrasted with the wrong conclusion 
$I \gg 1$ that would be obtained from the ``effective Hamiltonian'' 
(\ref{Heff}).
We have obtained the correct value of $I$ because 
we have taken account of the strong correlation between 
$\delta n_{ba}$ and $\delta n_{err}$. 
Since this correlation comes from 
the form of the gauge interaction, 
similar correlation should also be present
for any measuring apparatus of any gauge field.
Therefore, similar optimization of the apparatus
is necessary, and the response range and the information entropy 
would be finite, 
for any measuring apparatus of any gauge field.

\section*{Concluding remarks}

The above example clearly demonstrates that 
the limitation of available interactions in nature
gives rise to a fundamental limit of quantum  measurement.
I discuss possible consequences of this fundamental limit.

Clearly, it imposes a fundamental limit of getting information on 
quantum systems because the ``ideal measurement''
(i.e., the measurement with $\delta Q_{err} = \delta Q_{ba} = 0$,
although the backaction on the conjugate variable is of course
finite) is forbidden
for a certain set of observables such as 
the number of gauge quanta.
Namely, one cannot get information on $Q$ without disturbing
$Q$ itself.
If one wishes to make this backaction on $Q$ small, 
then 
the response range of the apparatus and the information entropy 
obtained by the measurement 
are limited, 
as demonstrated in the previous section.

It should be pointed out that 
this limitation of measurement leads to many 
other limitations.
For example, it imposes a limit on the 
controllability of quantum systems: 
One must perform a measurement to control a system,
but an accurate measurement gives rise to 
large backactions on some of the measured quantities, 
and the control becomes imperfect.
Another example is 
the limitation of the preparation of quantum states.
To prepare a system in a desired state,
one must perform ideal measurements on some set of 
observables, e.g., on a ``complete set of
commuting observables'' \cite{dirac}.
However, as I have shown in this paper, 
an ideal measurement is impossible for some
observables.
Therefore, the problems of the preparation 
of quantum states should be reconsidered.

As seen from these examples, the limitation of the measurement 
means that {\it accurate specification of 
quantum systems is impossible in general}.
This seems to be related to the foundation of 
the statistical mechanics.
For example, imperfect knowledge about the initial state of the 
system leads to rapid loss of knowledge as the time evolves. 
This results in the ergodic property and the mixing property 
\cite{dset}, although quantum systems cannot have these 
properties if the initial state is accurately known \cite{BB}.

\subsubsection*{Acknowledgment}
This work has been supported by the Core Research for 
Evolutional Science and Technology (CREST) of the Japan Science 
and Technology Corporation (JST).



\begin{references}
\bibitem{nonrel}
Actual calculations have been performed
using the nonrelativistic approximation to 
Eq.\ (\ref{qed}).
This can be justified easily.
\end{references}
\end{document}